\documentclass[conference]{IEEEtran}
\IEEEoverridecommandlockouts
\usepackage{cite}
\usepackage{amsmath,amssymb,amsfonts}
\usepackage{algorithmic}
\usepackage{graphicx}
\usepackage{textcomp}
\usepackage{xcolor}
\usepackage{flushend}
\def\BibTeX{{\rm B\kern-.05em{\sc i\kern-.025em b}\kern-.08em
    T\kern-.1667em\lower.7ex\hbox{E}\kern-.125emX}}
\begin{document}

\title{A Scalable Framework for Solving Fractional Diffusion Equations}

\author{\IEEEauthorblockN{Max Carlson}
\IEEEauthorblockA{\textit{University of Utah} \\
Salt Lake City, USA \\
mcarlson@cs.utah.edu}
\and
\IEEEauthorblockN{Robert M. Kirby}
\IEEEauthorblockA{\textit{University of Utah} \\
	Salt Lake City, USA \\
	kirby@cs.utah.edu}
\and
\IEEEauthorblockN{Hari Sundar}
\IEEEauthorblockA{\textit{University of Utah} \\
Salt Lake City, USA \\
hari@cs.utah.edu}
}

\maketitle

\begin{figure*}[h]
	\centerline{\includegraphics[width=\linewidth]{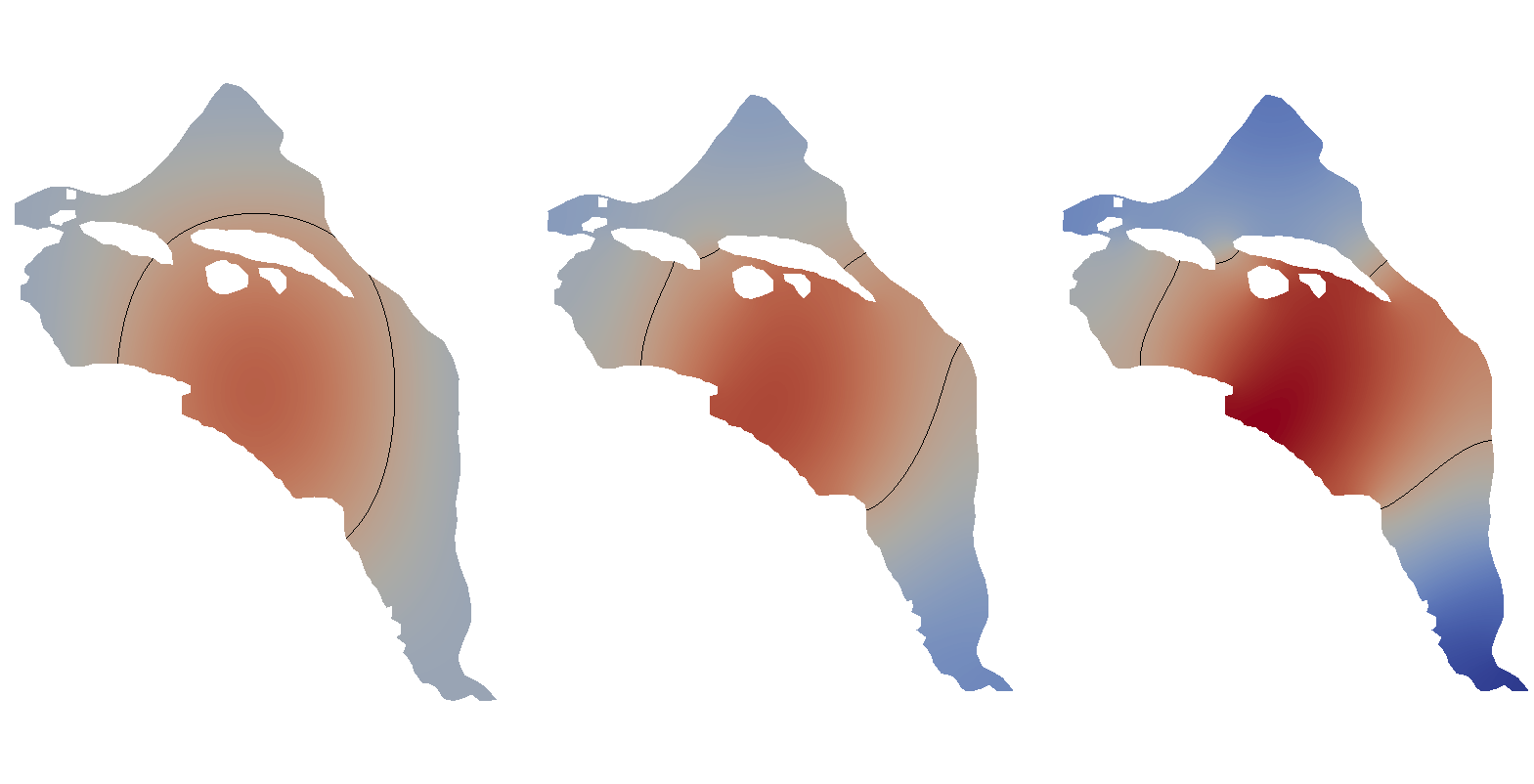}}
	\caption{Three example solutions to the fractional Poisson problem on complex geometry (Hanford site) with a polynomial forcing function. From left to right, the fractional order is $\alpha=0$, $\alpha=1$, and $\alpha=2$. When $\alpha=0$ we get the original forcing function and when $\alpha=2$ we get the standard Laplacian. The black line is the contour where the function value is equal to 0. By varying $\alpha$, we can smoothly interpolate between forcing function and its standard Laplacian.}
	\label{fig:title}
\end{figure*}

\begin{abstract}

The study of fractional order differential operators is receiving renewed attention in many scientific fields. In order to accommodate researchers doing work in these areas, there is a need for highly scalable numerical methods for solving partial differential equations that involve fractional order operators on complex geometries. These operators have desirable special properties that also change the computational considerations in such a way that undermines traditional methods and makes certain other approaches more appealing. We have developed a scalable framework for solving fractional diffusion equations using one such method, specifically the method of eigenfunction expansion. In this paper, we will discuss the specific parallelization strategies used to efficiently compute the full set of eigenvalues and eigenvectors for a discretized Laplace eigenvalue problem and apply them to construct approximate solutions to our fractional order model problems. Additionally, we demonstrate the performance of the method on the Frontera computing cluster and the accuracy of the method on simple geometries using known exact solutions.

\end{abstract}

%\begin{IEEEkeywords}
%component, formatting, style, styling, insert
%\end{IEEEkeywords}

\section{Introduction}
% introduction to the paper

% brief history of fractional operators

For almost as long as the concept of a derivative has existed, people have asked the question ``is there such a thing as a half derivative?'' Perhaps the first recorded example of this can be found in a written correspondence between mathematicians L'Hopital and Leibniz in 1695. L'Hopital asked Leibniz about his formula for the ``n-th'' derivative, specifically he asked about what would happen if you considered n = 1/2. At the time Leibniz did not have a concrete answer but mentioned it was an ``apparent paradox, from which one day useful consequences will be drawn'' \cite{Loverro2004FractionalC}.

% some key properties of fractional operators

Aside from intellectual curiosity, are these operators even useful? From an intuitional standpoint, what would a half derivative even mean? The first derivative has a straightforward interpretation as the instantaneous rate of change of some function but it is not immediately obvious what the interpretation would be for a fractional order derivative. The Riemann-Liouville and Caputo definitions \cite{definitions} of the fractional derivative give us a bit of insight here. These definitions evaluated a point $t$ are defined as an integral over the domain $(\infty, t]$ of some weight function multiplied by $f$ or $\frac{d^nf}{dt^n}$. This is where we can see the main and probably most important property that is different from integer order derivatives. Fractional order derivatives are \textbf{nonlocal}! Instead of being defined with respect to some infinitesimally small region of function values around $t$, the fractional derivative depends on every value of $f$ all the way from $-\infty$ up to $t$. What does this mean intuitively? Instead of describing an instantaneous rate of change, fractional derivatives describe a time-weighted rate of change where the order $\alpha$ essentially determines how much importance past function values have.

Differential equations with integer order operators are based on the assumption that the rate of change of the given process is completely independent on the state of the process in the past. This is a perfectly fine assumption for a lot of natural processes but it seems like a pretty big oversimplification to assume that this holds for any process in general.

% brief history of anomalous diffusion and connection to fractional laplacian

One example of how the usage of fractional operators expanded the understanding of a process comes from diffusion. \cite{Henry2010AnIT} \cite{10.3389/fphy.2019.00018} One of the key features of diffusion that comes from the integer order model is that the mean squared displacement scales proportionally with time. There have been experiments that show there are processes that look very much like diffusion but do not satisfy this condition. This type of diffusion process is known as anomalous diffusion and can be modeled using fractional time or space differential operators depending on the specific regime. For example, superdiffusion (one of the regimes of anomalous diffusion) can be modeled using the fractional Laplacian operator. 

For this paper, we are specifically interested in the fractional Laplacian operator \cite{lischke2018fractional}. It shows up in many models ranging from finance \cite{KUMAR2014177} \cite{Levendorskii2004PRICINGOT}, biology \cite{10.1371/journal.pone.0143938} \cite{MAGIN20101586} \cite{ANZIAMJ6283}, structural mechanics \cite{tarasov2018fractional} \cite{evgrafov2018nonlocal}, and quantum mechanics \cite{Laskin2002FractionalSE} \cite{GUERRERO2015604}. Most of these models are interested specifically in the 1D version of the fractional Laplacian or are limited to simple domains in higher dimensions. This may be partially due to the lack of widely available solvers for fractional PDEs on very large complex domains. 

% overview of applications of the fractional laplacian operator

% use all of this to justify the need for scalable, easy-to-use, fractional laplacian solvers to aid the research in this field

With all of this research happening in fractional PDEs, it is necessary to have scalable solvers so that approximate solutions to equations can be found in a reasonable amount of time so that scientists can verify and replicate results. We have developed one such solver and the following sections of this paper will go into detail on how our method works and the paralellization strategies employed to ensure we are taking full advantage of available computing resources.

% discuss why finite element method is less appealing in the fractional order case

\section{Method}
% spectral definition:

For this paper, we will use the spectral definition of the fractional Laplacian \cite{lischke2018fractional} specifically for the case of homogeneous boundary conditions (Dirichlet or Neumann). It is possible to extend this method to nonhomogeneous boundary conditions by using harmonic lifting but this is beyond the scope of this paper as it does not change the underlying principles of the method. The harmonic lifting can be computed using integer order solvers and does not change the need for homogeneous eigenpairs. This spectral definition is constructed by first considering the infinite, discrete set of eigenvalues and eigenfunctions of the integer order Laplace eigenvalue problem \eqref{eq:continuous_eigenproblem} with appropriate homogeneous boundary conditions.

\begin{equation}
\begin{array}{ccc}
-\Delta\psi_{k} & = & \lambda_{k}\psi_{k}\\
\mathcal{B}(\psi_{k})|_{\partial \Omega} & = & 0.
\end{array}
\label{eq:continuous_eigenproblem}
\end{equation}

With these eigenpairs, the spectral fractional Laplacian of a function $u$ is defined as

\begin{equation}
(-\Delta)^{\alpha/2}u=\sum_{k=1}^{\infty}\lambda_{k}^{\alpha/2}\langle u,\psi_{k}\rangle\psi_{k}
\label{eq:spectral_definition}
\end{equation}

\noindent where $\langle \cdot , \cdot\rangle$ is the inner product for $L_2(\Omega)$.

% model problem: fractional diffusion

The first model problem we will consider is the space-fractional diffusion equation \eqref{eq:model_diffusion} with homogeneous boundary conditions and some arbitrary initial conditions:

\begin{equation}
\begin{array}{ccc}
\partial_{t}u(x,t) & = & -\mu(-\Delta)^{\alpha/2}u\\
u(x,t)|_{\partial\Omega} & = & 0\\
u(x,0) & = & u_{0}(x).
\end{array}
\label{eq:model_diffusion}
\end{equation}

The nice thing about the spectral definition of the fractional Laplacian is that it can be plugged directly into the diffusion equation \eqref{eq:model_diffusion} and using linearity we obtain

\begin{equation}
\sum_{k=1}^{\infty}[\partial_{t}\hat{u}_{k}(t)+\mu\lambda_{k}^{\alpha/2}\hat{u}_{k}(t)]\psi_{k}=0
\label{eq:diffusion_intermediate}
\end{equation}

Notice that now we can solve this equation exactly by finding $\hat{u}_k(t)$ such that each eigenfunction $\psi_k$ is multiplied by 0. Specifically, we have transformed the problem into an infinite set of ODEs that can be each solved exactly. Therefore, this fractional diffusion problem can be solved directly using this method and gives the solution

\begin{equation}
u(x,t)=\sum_{k=1}^{\infty}e^{-\mu\lambda_{k}^{\alpha/2}t}\langle u_{0},\psi_{k}\rangle\psi_{k}.
\label{eq:diffusion_solution}
\end{equation}

% model problem: fractional poisson

Similarly, we consider the steady state of the diffusion equation with a forcing term. This equation \eqref{eq:model_poisson} is also known as the fractional Poisson equation with homogeneous boundary conditions:

\begin{equation}
\begin{array}{ccc}
(-\Delta)^{\alpha/2}u & = & f\\
u|_{\partial\Omega} & = & 0.
\end{array}
\label{eq:model_poisson}
\end{equation}

If $f$ is equivalent to an infinite sum of coefficients times the eigenfunctions that agree with our boundary conditions, 

\begin{equation}
f = \sum_{k=1}^{\infty}\langle f,\psi_{k}\rangle\psi_{k}
\label{eq:poisson_forcing}
\end{equation}

\noindent then we can once again directly plug the spectral definition and this $f$ into \eqref{eq:model_poisson}. Using the linearity of summations, the coefficients for the unknown function $u$ can be solved for directly to obtain

\begin{equation}
u = \sum_{k=1}^{\infty}\lambda_{k}^{-\alpha/2}\langle f,\psi_{k}\rangle\psi_{k}.
\label{eq:poisson_solution}
\end{equation}

For both of these model problems, we have not needed to discuss a numerical scheme for computing solutions. We have simply shown that both problems have an extensive family of exact solutions that contain the types of solutions we are looking for. For domains like lines, squares, and cubes, this eigenfunction expansion definition is simply a more general form of the Fourier series solution. 

% introduce method of eigenfunction expansion

This more general definition then also allows us to extend this approach to other complex geometries with known exact eigenfunctions like discs, spheres and cylinders. But what if we want to solve these problems on more interesting geometries that do not have known eigenfunctions? In this case, we will have to compute approximations to the eigenfunctions.

In order to turn this into a computable problem, we need to discretize the domain and using the weak form of the Laplace eigenvalue problem we obtain

\begin{equation}
K\phi_{k}=\theta_{k}M\phi_{k}
\label{eq:discrete_laplace}
\end{equation}

\noindent where $K$ is the stiffness matrix and $M$ is the mass matrix such that

\begin{equation}
K_{ij}=\langle\nabla e_{i},\nabla e_{j}\rangle\quad M_{ij}=\langle e_{i},e_{j}\rangle
\end{equation}

\noindent given basis functions $e_i$ and $e_j$.

Using the solution to this eigenvalue problem, the approximate solution is simply

\begin{equation}
\begin{array}{ccc}
g & = & \sum_{k=1}^{N}\langle f,\phi_{k}\rangle\phi_{k}\\
v & = & \sum_{k=1}^{N}\theta_{k}^{-\alpha/2}\langle f,\phi_{k}\rangle\phi_{k}
\end{array}
\end{equation}

where $g$ is the approximation of the forcing function using the computed eigenbasis and $v$ is the approximate solution to the fractional Poisson problem \eqref{eq:model_poisson}.

The important thing to note is that there is nothing fractional about this problem! This system \eqref{eq:discrete_laplace} is just the typical finite element discretization \cite{boffi_2010} of the Laplace eigenvalue problem and therefore the resulting $K$ and $M$ retain the expected sparsity. The tradeoff then is instead of solving an $N\times N$ dense system (fractional FEM), we solve for $N$ eigenpairs of the sparse, symmetric system \eqref{eq:discrete_laplace}. 

You might notice that both of these approaches (fractional FEM and eigenfunction expansion) result in the same algorithmic complexity so why should we solve for the discrete eigenfunctions? If you wanted to solve the fractional Poisson problem using finite element method, for each and every forcing function $f$, the dense $N\times N$ system would have to be solved. By using the approximate eigenfunctions, the eigenvalue problem only needs to be solved once and then can be reused for every new forcing function. Therefore the eigenfunction expansion approach results in more flexibility in the case of solving many problems for a single domain. Essentially, given the geometry and the basis function order, that is all that is needed to construct the fractional operator and that is the step that is most computationally expensive.

% justify need for computing all eigenpairs

\begin{figure}[h]
	\centerline{\includegraphics[width=\linewidth]{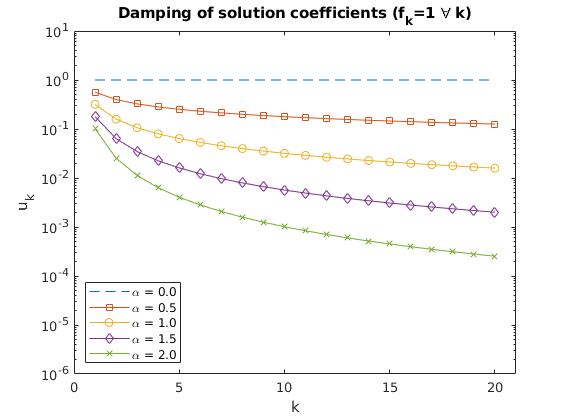}}
	\caption{Applying the fractional Laplacian can be viewed as a damping of the spectral coefficients. This example is what the damped coefficients look like for various $\alpha$ for the first 20 1D eigenvalues.}
	\label{fig:coeff_damping}
\end{figure}

\section{Implementation}
\subsection{Nektar++}

% why nektar++?

In order to ensure that our solver could be used by researchers without needing to change their typical workflow, we developed this solver to be fully integrated in the Nektar++ spectral/hp element framework \cite{CANTWELL2015205}. To solve a fractional diffusion or Poisson equation, the Nektar++ session files are exactly the same as for the integer order variants but now with an added $\alpha$ parameter. There are additional solver parameters that can be supplied but they have default values that work well in the majority of cases and will be detailed in the rest of the implementation section.

% solution pipeline (forward transform, assembly, solve, backward transform)

The exact operations for which Nektar++ is responsible is as follows. The first is the transformation of the input function into FEM expansion coefficients for whatever kind of basis you select. Then the stiffness and mass matrices are assembled according to the chosen basis as a sparse PETSc matrices. Then these matrices are given to our eigenvalue solver to compute the full set of eigenvalues and eigenvectors. Then, for any given forcing function, the eigenexpansion coefficients of the solution are computed and then given back to Nektar++ which finally does a backwards transform to go from FEM expansion coefficients back to actual function values evaluated typically at a collection of quadrature points. For a nodal basis, these forward and backwards transforms are just the identity since a nodal basis is designed to have coefficients equal to function values at given points.

The main operation that dominates the runtime of this solver is then solving for all eigenpairs of the discrete system. Since we are solving for all $N$ eigenpairs of a sparse system, the time complexity is $O(N^2)$ which is not ideal. Therefore we needed a parallelization strategy that reduces this amount of work as much as possible and takes full advantage of available hardware. 

\subsection{Solving for all eigenpairs in an interval}

Before we go into further details of our method, some background is necessary to understand how eigenvalue problems are typically solved. One family of algorithms for this problem are Krylov subspace methods. The idea is to produce a tridiagonal matrix iteratively that has eigenvalues and eigenvectors that are approximately equal to the true eigenpairs. At each iteration, both the number of columns and rows of the tridiagonal matrix increases by one, which also means the number of approximate eigenvalues increases by one. 

This kind of iterative scheme is usually terminated when some predetermined number of approximate eigenvalues have converged to the true values. For these methods, the eigenvalues that converge are those that are closest to 0. This idea can then be extended by solving for eigenvalues of the shifted system $(K-aM)$. The true eigenvalues of this shifted system close to 0 can be shifted back to get the true eigenvalues of the original system that are closest to some arbitrary point $a$. With this, we can iteratively solve for the $d$ closest eigenpairs to any given value $a$. 

Even better, solving for the $d$ closest eigenvalues to $a$ and the $d$ closets eigenvalues to another point $b$ can done completely indepedently of each other. If the total set of eigenvalues can be partitioned into groups of $d$ eigenpairs, then we can use SLEPc's Krylov-Schur iterative method to compute each batch of eigenpairs completely indepedently. SLEPc \cite{Hernandez:2005:SSF:1089014.1089019} is a library for solving eigenvalue problems that is built on top of the parallel linear algebra library PETSc \cite{petsc-user-ref}. This forms the first level of parallelism for our approach and is typically known as spectrum slicing \cite{li2018eigenvalues}. The tricky part is then how to split the interval containing all solutions into load-balanced subproblems. This is the main focus of our approach and will be elaborated in more detail a bit later.

% SLEPc
% Krylov-Schur method
% Cholesky factorization for system solves

\subsection{Two-level parallelism and communication hierarchy}

Before we get into how we do the partitioning, we need to go over the overall parallelization strategy. The first layer of parallelization is clear, if we have $P$ processors, and we are solving for $N$ eigenvalues, then each processor is indepedently solving for $\frac{N}{P}$ eigenpairs. This approach is not ideal since this would mean each processor has to solve systems of the form $K-aM$ without taking advantage of any parallelism.

If we instead have $P$ groups of $p$ processors, then all $p$ processes can solve their shifted systems in parallel. We will refer to each group of processors as an ``evaluator''. Typically, an evaluator will simply be a single machine but it doesn't necessarily have to be. there can be multiple evaluators per physical machine or even multiple physical machines per evaluator.

\subsection{Partitioning}

At last we come to the main focus of our approach, how to partition the total set of eigenpairs into equally sized, contiguous groups of eigenpairs that can be solved for independently. For any given system, the spectral radius $\rho$ can be approximated with a few applications of $K$ and $M$. Since the system is symmetric, we know all possible eigenvalues are real and lie in the interval $[-\rho, \rho]$. For the standard Laplace problem, we also have positive semidefiniteness so this can be taken even further to say all eigenvalues are in $[0,\rho]$. In order to partition this interval into subintervals with equal number of eigenpairs, we need some way to count the number of eigenvalues in a subinterval.

\subsubsection{Eigenvalue counting}

There are then two approaches we considered for the problem of eigenvalue counting \cite{napoli2013efficient}. The first is an approximate technique known as kernel polynomial filtering that only requires matrix-vector multiplications and the second is a more expensive ``exact'' technique that requires matrix factorization.

% approximate approach

The kernel polynomial filtering approach uses the Chebyshev series approximation of the heaviside step function to approximate the eigenprojector of $A=K-aM$, $A^{\star}$:

\begin{equation}
A^{\star}=\sum_{k=1}^{K}\gamma_{k}T_{k}(K-aM)
\end{equation}

Essentially, this method is creating a new matrix that has only the eigenpairs greater than or equal to $a$. Then the number of eigenvalues greater than $a$ can be computed as the trace of this new approximate matrix $A^\star$:

\begin{equation}
\text{tr}(A^{\star})\approx\mathcal{N}(\lambda\geq a).
\end{equation}

The beauty of this approach is that in order to evaluate $T_{k}$ and $A^\star$, the only operations required are matrix addition and matrix-vector multiplication. Since the system we are interested in is sparse, this is even more ideal since these operations are like $O(N)$. Unfortunately, this approach is not very stable since the approximation of the step function does not satisfy certain constraints. Specifically, the filter can be negative and isn't monotonically increasing. Therefore the number of terms you need to get even remotely consistent or accurate counts is so high that it takes longer than the exact method. This approach is okay if you only need rough counts but for partitioning, we need very accurate counts.

% exact approach

Therefore, we use an ``exact'' counting technique that relies on Cholesky factorization. Specifically, using Sylvester's Law of Inertia, we get that the number of eigenvalues of $A=K-aM=LDL^{T}$ greater than or equal to 0 is the same as the number of entries of the diagonal matrix $D$ greater than or equal to 0 where $D$ is the diagonal matrix resulting from the Cholesky factorization. This is more expensive than simple matrix-vector multiplications but gives consistent and accurate results in all cases.

Equipped with a way to count the number of eigenvalues in an interval $[a,b]$, we can now partition the total interval $[0,\rho]$ into equally sized subintervals. So far, none of the things mentioned are new, in fact, SLEPc's Krylov-Schur solver is designed to accommodate the spectrum slicing approach. However, SLEPc has no methods for actually forming the subintervals and if the partitioning is not supplied, it will use evenly spaced subintervals which result in massive load imbalance even on simple domains. Therefore we had to develop our own partitioning scheme to get proper load balancing.

\subsubsection{Tree partitioning}

\begin{figure}[h]
	\centerline{\includegraphics[width=\linewidth]{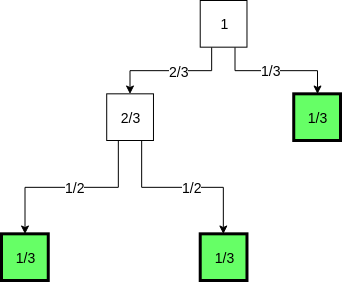}}
	\caption{Tree based partitioning. In order to split all eigenvalues into three equal-sized groups, first split the total interval into 2/3 and 1/3 sized groups and then split the 2/3 sized group in half. Each split involves a binary search style operation where each step requires a Cholesky factorization.}
	\label{fig:tree_partition}
\end{figure}

The first approach we tried was tree partitioning. With the ability to count the number of eigenvalues greater than some point $a$, we could do a binary search to split the total interval into two subintervals of arbitrary size. Then for any given $P$, the interval could be split in a tree-like fashion (an example can be seen in Figure \ref{fig:tree_partition}) to get evenly loaded subintervals. Unfortunately, this approach had a number of problems that dramatically hurt scalability. 

The total amount of work required to do the partitioning using this scheme is

\begin{equation}
\tau_P \approx \log_2(P) n_c T(k,N,p),
\end{equation}

\noindent where $P$ is the number of evaluators, $p$ is the number of processors per evaluator, $k$ is some measure of the sparsity of the system, $n_c$ is the number of iterations the binary search takes to get a good split, and $T$ is the time it takes to do the parallel Cholesky factorization of the shifted system $K-aM$.

The problem here should be clear, every time we add another machine, the partitioning time grows! Even if $log_2(P)$ grows slowly with $P$, the $n_c T(k,N,p)$ portion is pretty expensive so multiplying it by some scaling factor is not great. What we really want is something that is constant with respect to $P$. Another big issue is that at any given step of the tree partitioning, most of the evaluators are doing absolutely nothing. Even in the best case, at any level, at most $P/2$ evaluators are doing anything.

\subsubsection{Greedy partitioning}

Not only did we want to avoid scaling the partitioning time with the number of evaluators, but if possible, avoid doing any binary searches as well. Each binary search requires around 10 Cholesky factorizations in order to get a good split which adds up really fast. Additionally, all of those 10 steps have to be done by a single evaluator and cannot really be aided by any idle machines. Therefore we came up with a greedy partitioning strategy that works as follows.

We can get an initial guess for the partitioning by splitting the total interval into $P$ subintervals of equal length. This is actually the default behavior of SLEPc if no partitions are provided to a spectrum slicing method. This initial guess is actually pretty bad even for simple geometries since eigenvalues are not evenly distributed through the spectrum. 

\begin{figure*}[h]
	\centerline{\includegraphics[width=\linewidth]{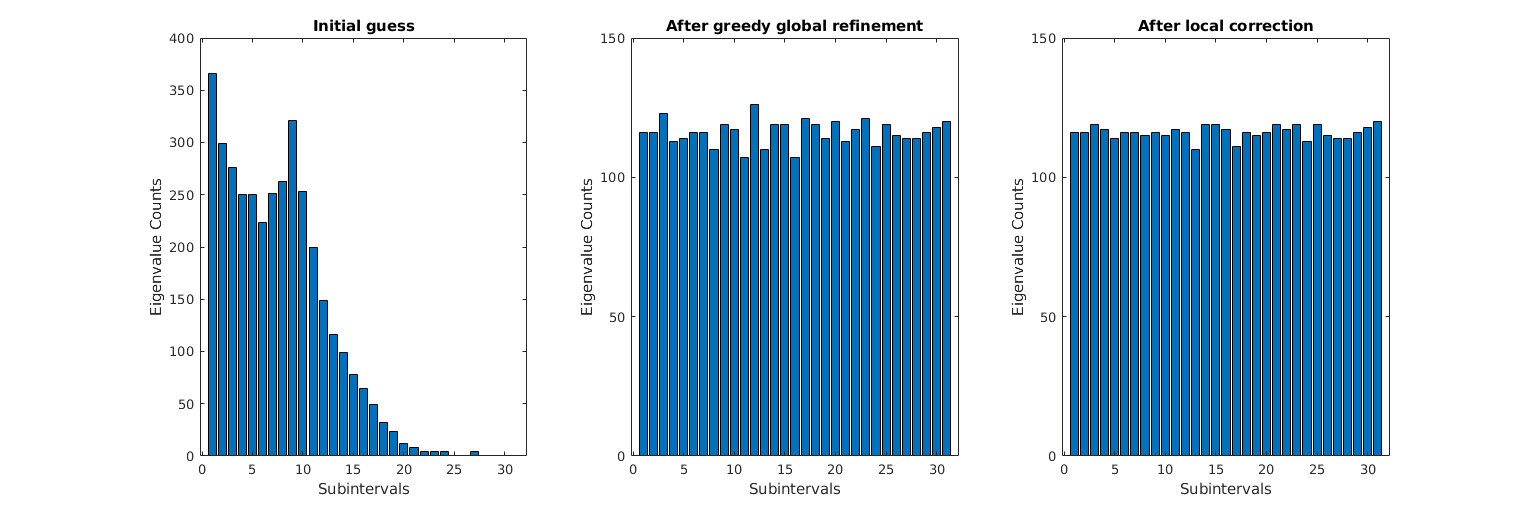}}
	\caption{Greedy partition refinement. The initial guess of equally space subintervals (default behavior in SLEPc) results in unbalanced partitions. Greedy global refinement gets close to ideal and then local refinement fixes any especially bad partition pairs.}
	\label{fig:partition_stages}
\end{figure*}

With this initial guess, all but one evaluators each compute the number of eigenvalues greater than one of the splitting points in a single Cholesky factorization in parallel. With these values, the number of eigenvalues in each subinterval can be computed as can be seen in the leftmost plot in Figure \ref{fig:partition_stages}. This histogram can be thought of as a piecewise constant approximation to the true density of states for the system. Then the root processor uses this approximation of the density to choose new splits that would give equally loaded subintervals if this was the true density and then this process repeats $n_a$ times where $n_a$ is some user supplied parameter.

Unfortunately, this greedy iterative approach does not converge to the true optimal partitioning. Instead it gets very close to optimal and then enters a cycle. However, the best partitioning can be stored at each step and is reached usually within 3 or 4 iterations. This greedy global refinement only takes a handful ($n_a$ has a default of 7) of Cholesky factorization and provides a partitioning that is much closer to the optimal than the naive initial guess.

In certain rare cases, the best partitioning produced by the global refinement can have pairs of partitions that are very far from the optimal. For these cases, a second optional local refinement stage can be employed. In order to do this in such a way that does not scale with $P$, for each iteration, half of the evaluators do a binary search on their pair of subintervals (if they are unbalanced beyond a given threshold) to balance the pair towards the optimal and then other half of evaluators balance the remaining pairs. This can then be done $n_b$ times but typically one pass handles any major outliers.

With this strategy, the total time complexity for partitioning is

\begin{equation}
\tau_P \approx (n_a + 2n_b n_c)T(k,N,p)
\end{equation}

\noindent where $n_a$ is the number of global refinement steps (default 7), $n_b$ is the number of local refinement passes (default 3), and $n_c$ is the number of iterations required to get a good split by a binary search (default 10). This is exactly the kind of time complexity we are looking for. The number of evaluators $P$ does not show up at all and the total number of Cholesky factorizations is completely determined by user-supplied constants. It should also be noted that this is a worst case complexity. In practice, the local refinement is rarely triggered and even then usually terminates in a single pass.

\subsection{Post-processing}

% symmetry of problem ensures orthogonalization can be done entirely local to each node independently

Finally, once the complete set of eigenpairs are computed, they need to be orthogonalized and normalized to ensure the eigenvectors are orthonormal. Fortunately, since the system is symmetric, the only eigenvectors that need to be made orthogonal to each other are those that share the same eigenvalue. Since any eigenvectors that share an eigenvalue will fall into the same subinterval, the very same evaluator that computed these vectors can do the orthogonalization completely independently from any other evaluator.

\section{Performance}
The overall time complexity using our partitioning algorithm is then

\begin{equation}
\tau \approx \frac{N}{P}T(k,N,p) + (n_a + n_b n_c)T(k,N,p) + R
\end{equation}

\noindent where the first term is the eigenvalue problem solve phase, the second term is the partitioning phase, and the last term, R, is the remainder. The remainder consists of the time taken by Nektar++ to assemble the discretized system and for some other small and quick operations like computing the spectral radius. This time is just left as a remainder since it is completely dominated by the solve and partitioning phases. Another note is the time to solve for a single eigenpair is dominated by the time it takes to do a single parallel Cholesky factorization, therefore solving for $\frac{N}{P}$ eigenpairs takes about $\frac{N}{P}T(k,N,p)$ time.

To test out the performance and accuracy for the solver, we ran experiments primarily on two computing clusters. The first is the University of Utah CHPC's Kingspeak cluster which accommodated jobs up to 12 nodes where each node is dual socket with Intel Xeon processors and 64GB of memory. For larger jobs, we used the Texas Advanced Computing Center's new Frontera computing system during its early access phase. Frontera is a powerful new cluster with 8,008 nodes with Intel Platinum 8280 processors (also dual socket) and 192 GB of memory per node. Utilizing this cluster allowed us to push the size of the jobs to a more extreme scale utilizing hundreds to thousands of nodes.

\subsection{Test Meshes}

\begin{figure}[h]
	\centerline{\includegraphics[width=\linewidth]{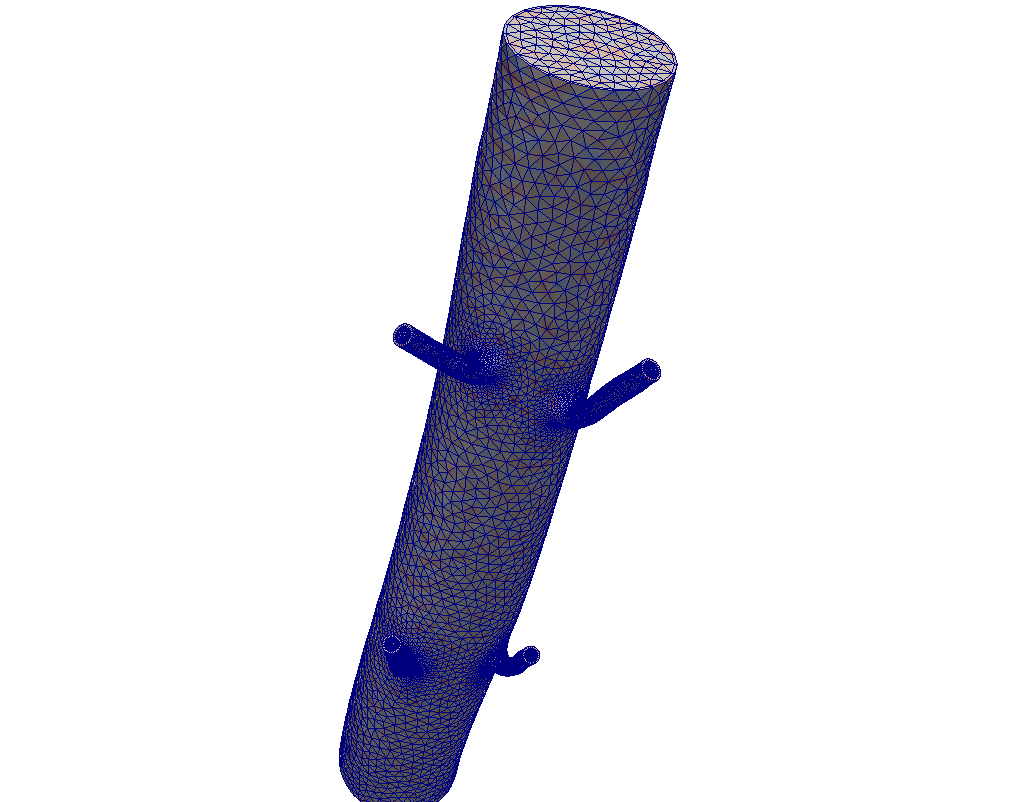}}
	\caption{Mesh of an aorta with about 24k tetrahedral elements.}
	\label{fig:aorta}
\end{figure}

\begin{figure}[h]
	\centerline{\includegraphics[width=\linewidth]{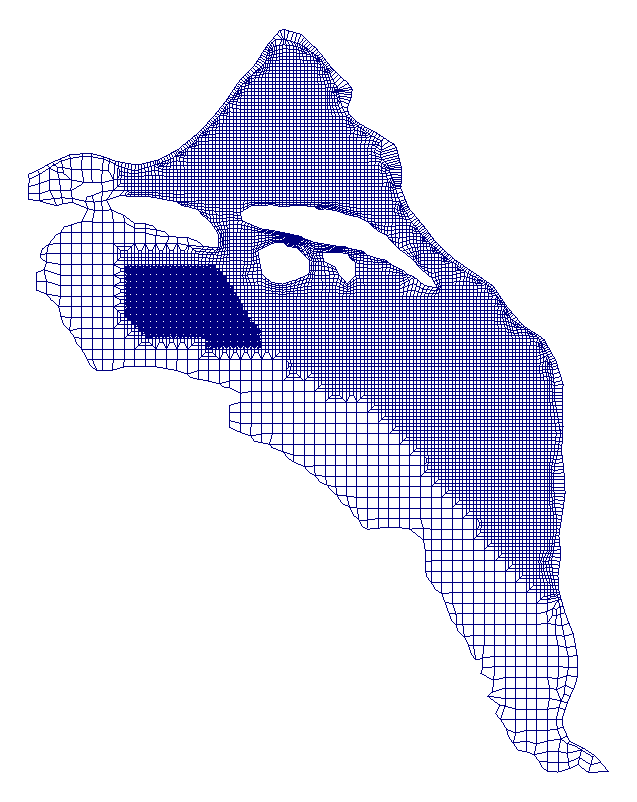}}
	\caption{Mesh of the Hanford site with about 15k quadrilateral elements.}
	\label{fig:hanford_mesh}
\end{figure}

To test how the solver works on complex geometry, we used two test meshes, one in 2D and one in 3D. For the 2D mesh, we used a mesh of the Hanford site which can be seen in Figure \ref{fig:hanford_mesh} which has about 15k quadrilateral elements. For the 3D mesh, we used a mesh of an aorta that can be seen in Figure \ref{fig:aorta} which has about 24k tetrahedral elements. The number of degrees of freedom for both of these meshes can be increased by increasing the order of the basis elements in order to simulate larger and higher order examples.

\subsection{Parallel Cholesky Factorization}

\begin{figure}[h]
	\centerline{\includegraphics[width=\linewidth]{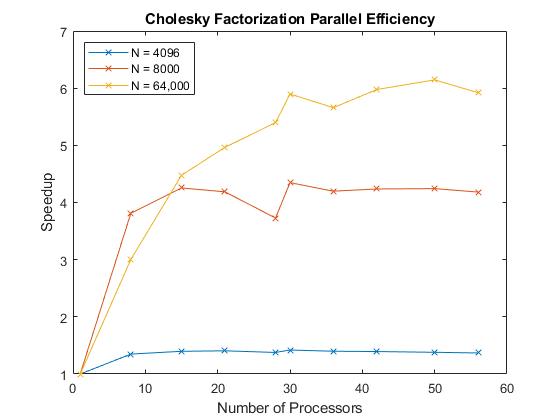}}
	\caption{Parallel speedup of Cholesky factorization for various $N$ and $p$. Run on the Kingspeak cluster with 28 processors per node.}
	\label{fig:cholesky}
\end{figure}

Before we get into the solver performance, there is still the question of how to determine the configuration parameters of the two layers of parallelism, $p$ and $P$. Specifically, how does the value of $p$ affect the time it takes to do a parallel Cholesky factorization? For this, we selected a few simple sizes of cubes and factored the matrix $K-aM$ and measured the time taken for each value $p$. This experiment was run on the Kingspeak computing cluster and the results (plotted as parallel speedup) can be seen in figure \ref{fig:cholesky}. Additionally, we use the MUMPS \cite{MUMPS:1} parallel Cholesky factorization library.

From this, it can be seen that the speedup from increasing $p$ reaches diminishing returns fairly rapidly and is dependent on the size of the system. For very small problems, it makes sense to have many evaluators per node since the speedup maxes out at something like 8 processors. However, our solver is geared towards very large scale problems and for these it makes more sense to use the full set of processors on a node. The default value then for $p$ in our solver is to use all of the processors in a socket. Otherwise an optimal pair of $P$ and $p$ need to be computed for the given problem.

\subsection{Strong Scalability}

\begin{figure}[h]
	\centerline{\includegraphics[width=\linewidth]{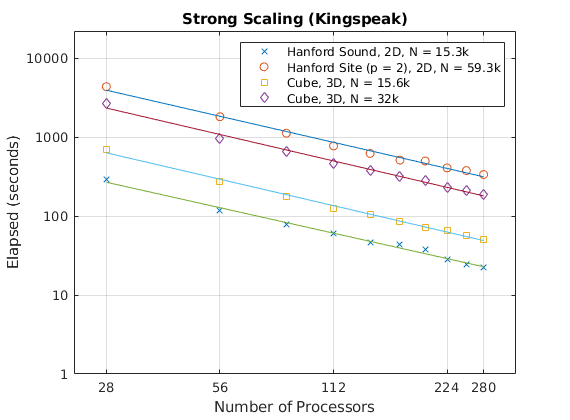}}
	\caption{Strong scaling for a collection of simple and complex 2D and 3D meshes on the Kingspeak computing cluster.}
	\label{fig:strong_scaling}
\end{figure}

In order to verify that the solver phase actually scales like $\frac{N}{P}T(k,N,p)$ we ran a strong scaling experiment on a variety of 2D and 3D meshes with both simple and complex geometry with different numbers of degrees of freedom. The results of this can be seen in Figure \ref{fig:strong_scaling}. The left plot shows the actual elapsed time with respect to $P$ and then the right plot shows the parallel speedup vs the ideal parallel speedup.

\begin{figure}[h]
	\centerline{\includegraphics[width=\linewidth]{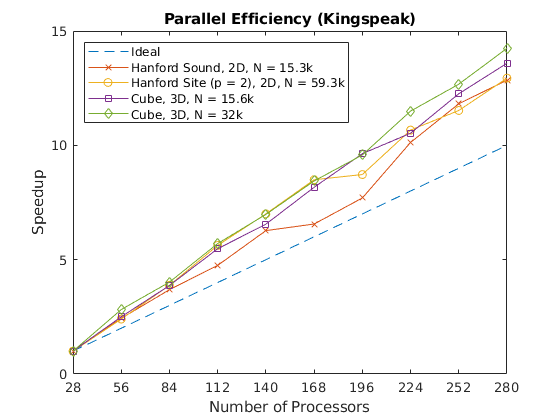}}
	\caption{Parallel speedup for a collection of simple and complex 2D and 3D meshes on the Kingspeak computing cluster.}
	\label{fig:strong_scaling_eff}
\end{figure}

For these small values of $P$, we actually see something quite nice. The parallel speedup is greater than the ideal speedup in all cases. This is because the Krylov-Schur algorithm requires quite a lot more memory to compute $N$ eigenpairs than $N/2$ and this memory requirement must exceed what can be cached so there is excessive memory movement for the $P=1$ case that is mitigated when the problem is partitioned with larger $P$.

\begin{figure}[h]
	\centerline{\includegraphics[width=\linewidth]{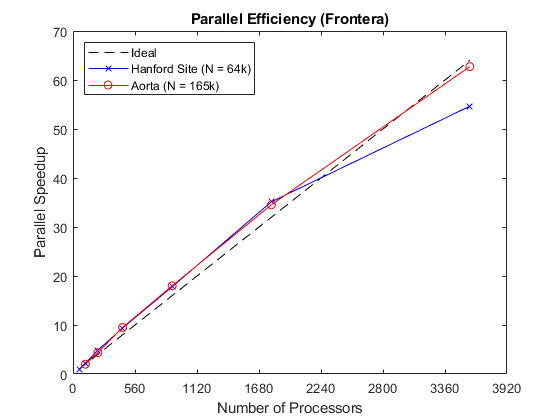}}
	\caption{Parallel speedup for Hanford site mesh and the aorta mesh on the Frontera computing cluster for P up to 64 nodes. A single node didn't have enough memory for the aorta so the baseline for the aorta mesh is $P = 2$. Parallel efficiency can be seen to drop off in the Hanford site once the noise in the greedy partitioning scheme dominates the magnitude of the optimal partitioning.}
	\label{fig:strong_scaling_frontera}
\end{figure}

To see what happens for larger $P$, we recreated this experiment on the Frontera cluster on a larger complex 3D mesh. The results of this can be seen in Figure \ref{fig:strong_scaling_frontera}. Eventually, as $P$ keeps increasing, the parallel speedup does start to dip under the ideal linear speedup. In the case of the Hanford site mesh, as $P$ goes from 32 to 64, the parallel speedup drops off dramatically. The greedy partitioning scheme doesn't produce the optimal partitioning but instead an approximation of it with noise. At a certain point, the magnitude of this noise outweighs the magnitude of the optimal solution and there will be sub-intervals with too many eigenpairs that end up inflating the total execution time.

\subsection{Weak Scaling}

With the scalability behaving as expected from the time complexity estimate, the next step was to see how far the solver can be pushed. Since the key idea motivating this work is that if we need to compute something like a million eigenpairs, we wanted to be able to throw more and more machines at the problem until we could compute them all in some small amount of time on the order of a few minutes. In order to see how many eigenpairs we could actually get with a realistic number of machines, we set up a range of problem sizes from 24k eigenpairs to half a million and increased $P$ until we got a solve time under 10 minutes. This experiment was done on the Frontera cluster and the raw data can be seen in Table \ref{tab:weak}. 

\begin{figure}[h]
	\centerline{\includegraphics[width=\linewidth]{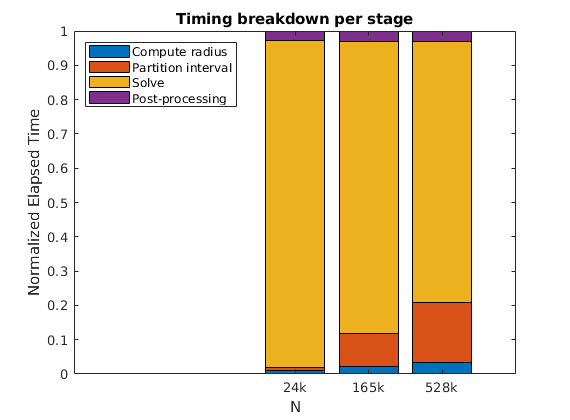}}
	\caption{Normalized timing for each stage of the solver for the aorta mesh with three different orders of basis functions. $N$ is the total degrees of freedom and the number of eigenpairs for which we are solving.}
	\label{fig:weak_scaling}
\end{figure}

Additionally, in Figure \ref{fig:weak_scaling}, the normalized run times can be seen to get an idea of how much of the total run time is being spent in each phase. The solve phase (plotted as the yellow bar) is the portion of the runtime that can be reduced by increasing $P$. Computing the spectral radius and the post-processing phase are staying roughly constant (as a percentage of total runtime) as $N$ increases but the partition time increases with $N$.

\begin{table}[h]
	\caption{Raw data for aorta weak scaling experiment.}
	\begin{center}
		\begin{tabular}{|c|c|c|c|}
			\hline 
			\textbf{Number of DoFs} & \textbf{24k} & \textbf{165k} & \textbf{528k} \\ 
			\hline 
			&  &  &  \\ 
			\hline 
			Number of Nodes & 1 & 64 & 512 \\ 
			\hline 
			Number of Evaluators & 2 & 128 & 512 \\ 
			\hline 
			Threads per Evaluator & 28 & 28 & 56 \\ 
			\hline 
			Total Processes & 56 & 3,584 & 28,672 \\ 
			\hline 
			&  &  &  \\ 
			\hline 
			\textbf{Total Elapsed} & \textbf{4m59s} & \textbf{4m23s} & \textbf{9m33s} \\ 
			\hline 
			(Compute spectral radius) & 3s & 6s & 20s \\ 
			\hline 
			(Partition interval) & 3s & 25s & 1m40s \\ 
			\hline 
			(Solve) & 4m44s & 3m44s & 7m14s \\ 
			\hline 
			(Post-processing) & 8s & 8s & 18s \\ 
			\hline 
		\end{tabular} 
		\label{tab:weak}
	\end{center}
\end{table}

\section{Accuracy}
The previous section demonstrated the performance of our implementation but lacks context. For instance, how many actually eigenpairs do we need to get good accuracy? Is it overkill or is it not enough? In order to make this concrete, we also run a series of accuracy experiments to obtain this needed information.

To quantify the accuracy, we use the two model problems mentioned in the previous sections. The first is the diffusion equation with no forcing term. This represents the behavior of the homogenous solution to a more general diffusion equation. The second model problem represents the steady state solution as the homogeneous solution goes to 0. By separating the general diffusion equation into these two situations we can get a better idea of how the number of elements and basis order affects each regime in isolation.

The easiest way to test the accuracy is to use a simple domain like a square or a cube with known eigenfunctions that are easy to compute. For a square and cube, these are just the typical fourier modes. For these experiments, we also use homogeneous Dirichlet boundary conditions.

For the fractional homogeneous diffusion problem, we used the following initial conditions

\begin{equation}
u_{0} = 2\sin(\pi x)\sin(\pi y)
\label{eq:diffusion_ics}
\end{equation}

\noindent which then have the exact solutions

\begin{equation}
u = 2e^{-\mu(2\pi^{2})^{\alpha/2}t}\sin(\pi x)\sin(\pi y).
\label{eq:diffusion_sols}
\end{equation}

For the fractional Poisson problem, we used the same functions for the forcing,

\begin{equation}
f = 2\sin(\pi x)\sin(\pi y)
\label{eq:poisson_f}
\end{equation}

\noindent which have exact solutions

\begin{equation}
u = 2(2\pi^{2})^{-\alpha/2}\sin(\pi x)\sin(\pi y).
\label{eq:poisson_u}
\end{equation}

Using these exact solutions, we computed approximate solutions for a variety of inputs. The first input is the order of the basis functions and we varied this value from 1 to 8. The second input is the number of elements in the mesh. For each of these, the solution error was computed for 11 equally spaced values of $\alpha$ from 0 to 2. The solution error for the diffusion equation was evaluated at $t = 0.4$ and $\mu = 1$.

\begin{figure}[h]
	\centerline{\includegraphics[width=\linewidth]{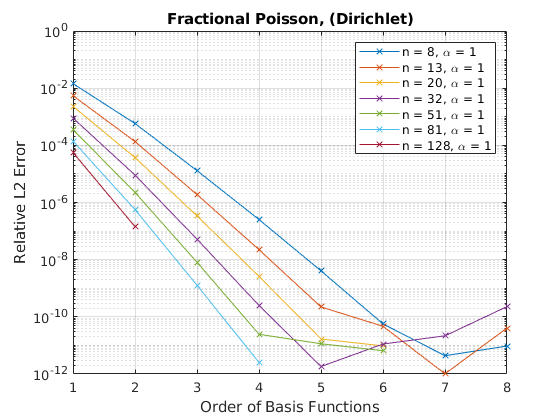}}
	\caption{Solution accuracy for fractional Poisson equation with respect to order of basis functions and number of elements.}
	\label{fig:accuracy_poisson}
\end{figure}

\begin{figure}[h]
	\centerline{\includegraphics[width=\linewidth]{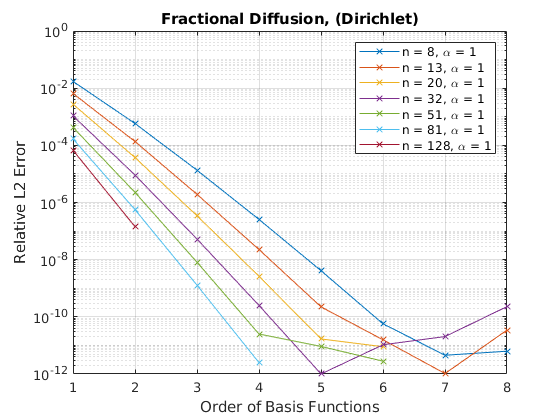}}
	\caption{Solution accuracy for fractional diffusion equation with respect to order of basis functions and number of elements.}
	\label{fig:accuracy_diffusion}
\end{figure}

The results of this equation for the Poisson equation can be seen in Figure \ref{fig:accuracy_poisson} and the results for homogeneous diffusion can be seen in Figure \ref{fig:accuracy_diffusion}. The main thing to note is that solution accuracy improves exponentially with respect to the order of the basis function. Increasing the number of elements improves the solution accuracy but not quite as drastically. This is because increasing the order improves accuracy at the low end of the spectrum where the input function ``lives'' but an input that has components in the high end of the spectrum might suffer with increased order.

While these plots show exponential convergence with respect to the order of the basis functions, the input functions are overly simple. Specifically, they are the first eigenfunctions of the Laplace operator and the approximation will reside almost entirely in the low end of the spectrum. Inputs that span more of the spectrum will likely require higher orders and number of elements in order to achieve the same level of accuracy.

\section{Conclusion}
We have developed a solver for fractional diffusion problems that can scale to any number of machines and that is fully integrated into Nektar++. The result is that now fractional diffusion can be simulated on very large complex geometries in a reasonable amount of time. Since our solver is integrated into Nektar++ with a mostly identical interface as typical diffusion problems, there should be no barrier to entry for scientists to immediately use this framework to test their ideas at a much larger scale. We hope that when this solver is pushed into the release branch of Nektar++ that it will be the first step in facilitating even more research into fractional operators. As a reminder, Nektar++ was used as a proxy for a general FEM solver and the eigenvalue solver itself does not require Nektar++ and can be applied broadly.

Now that the framework is up and running for homogeneous boundary conditions, the next step is to include support for nonhomogeneous BCs. According to \cite{lischke2018fractional}, there are a few approaches to handling these kinds of boundary conditions and further work will be in evaluating and implementing these methods. 

Additionally, the largest performance bottleneck at the moment is the time it takes to solve each shifted ($K-aM$) system using parallel Cholesky factorization. Since these systems are very sparse, we would like to utilize a preconditioned iterative method to approximate the solutions to these systems in a much smaller amount of time. One approach would be to use a multigrid based solver which would hopefully have the kind of convergence needed to beat a direct method like Cholesky factorization. We are looking into one such library that was developed for use with Nektar++ and further details of which can be found in \cite{8547580}.

The second performance bottleneck is the fact that we have to rely on exact eigenvalue counting techniques to get accurate enough counts for partitioning. The approximate technique is simply too unstable since the Chebyshev series representation of the step function doesn't satisfy the nonnegativity or monotonicty constraints necessary to accurately filter the spectrum. With a more stable approximate counting technique, the performance of the partitioning phase could be also improved by avoiding the need for parallel Cholesky factorization and relying solely on matrix vector multiplications.

\section*{Acknowledgment}

The first and third authors acknowledge the support of NSF (under DMS-1521748) and ARO (MURI 00001271 subcontract with Brown University) respectively.  The second authors acknowledges support from the Army Research Laboratory (ARL) under Cooperative Agreement Number W911NF-12-2-0023. The views and conclusions contained in this document are those of the authors and should not be interpreted as representing the official policies, either expressed or implied, of ARL or the US Government. The US Government is authorized to reproduce and distribute reprints for Government purposes notwithstanding any copyright notation herein.

\bibliographystyle{unsrt}
\bibliography{refs}

\end{document}